# Multi-Object Spectroscopy with the European ELT: Scientific synergies between EAGLE & EVE

C. J. Evans[1], B. Barbuy[2], P. Bonifacio[3], F. Chemla[3], J.-G. Cuby[4], G. B. Dalton[5,6], B. Davies[7], K. Disseau[3], K. Dohlen[4], H. Flores[3], E. Gendron[8], I. Guinouard[3], F. Hammer[3], P. Hastings[1], D. Horville[3], P. Jagourel[3], L. Kaper[9], P. Laporte[3], D. Lee[1], S. L. Morris[10], T. Morris[10], R. Myers[10], R. Navarro[11], P. Parr-Burman[1], P. Petitjean[12], M. Puech[3], E. Rollinde[12], G. Rousset[8], H. Schnetler[1], N. Welikala[13], M. Wells[1], Y. Yang[3,14]

[1] UK Astronomy Technology Centre, Royal Observatory Edinburgh, Blackford Hill, Edinburgh, EH9 3HJ, UK
[2] Universidade de São Paulo, IAG, Rua do Matão 1226, Cidade Universitária, São Paulo, 05508-900
[3] GEPI, Observatoire de Paris, CNRS, Univ. Paris Diderot, Place Jules Janssen, 92190 Meudon, France
[4] Aix Marseille Université, CNRS, LAM UMR 7326, 13388, Marseille, France
[5] Astrophysics, Department of Physics, Keble Road, Oxford OX1 3RH
[6] Space Science and Technology, Rutherford Appleton Laboratory, HSIC, Didcot OX11 0QX
[7] Institute of Astronomy, University of Cambridge, Madingley Road, Cambridge, CB3 0HA, UK
[8] LESIA, Observatoire de Paris, CNRS, Univ. Paris Diderot, Univ. P. & M. Curie, Place J. Janssen, 92195 Meudon, France
[9] Astronomical Institute Anton Pannekoek, Amsterdam University, Science Park 904, 1098 XH, Amsterdam, The Netherlands
[10] Department of Physics, Durham University, South Road, Durham, DH1 3LE, UK
[11] NOVA ASTRON, PO Box 2, 7990 AA Dwingeloo, The Netherlands
[12] Institut d'Astrophysique de Paris, 98bis Blvd. Arago, 75014 Paris, France
[13] Institut d'Astrophysique Spatiale, CNRS & Univ. Paris Sud XI, 91405 Orsay Cedex, France
[14] National Astronomical Observatories, Chinese Academy of Sciences, Beijing, China

**ABSTRACT**

The EAGLE and EVE Phase A studies for instruments for the European Extremely Large Telescope (E-ELT) originated from related top-level scientific questions, but employed different (yet complementary) methods to deliver the required observations. We re-examine the motivations for a multi-object spectrograph (MOS) on the E-ELT and present a unified set of requirements for a versatile instrument. Such a MOS would exploit the excellent spatial resolution in the near-infrared envisaged for EAGLE, combined with aspects of the spectral coverage and large multiplex of EVE. We briefly discuss the top-level systems which could satisfy these requirements in a single instrument at one of the Nasmyth foci of the E-ELT.

**Keywords:** instrumentation: adaptive optics; ELTs; spectrographs – galaxies: evolution; stellar content

## 1. INTRODUCTION

The workhorse instruments of the 8-10m class observatories have become their multi-object spectrographs (MOS), providing comprehensive follow-up to both ground-based and space-borne imaging. With the advent of deeper imaging surveys from, e.g., the *HST* and VISTA, there are already a plethora of spectroscopic targets beyond the sensitivity limits of current facilities; this will continue to grow even more rapidly, e.g., after the completion of ALMA and the launch of the *JWST*. Therefore, one of the key requirements underlying plans for the next generation of ground-based telescopes, the Extremely Large Telescopes (ELTs), is for even greater sensitivity for optical and infrared (IR) spectroscopy. Indeed, with only three ELTs planned (for now) and their large construction/operations costs, the need to make efficient use of their focal planes becomes even more compelling than at current facilities.

In parallel to the Phase B design of the European Extremely Large Telescope (E-ELT; e.g. Gilmozzi & Spyromilio, 2008), nine Phase A studies were undertaken for potential instruments (for a full summary see Ramsay et al. 2010). The instrument studies spanned a vast range of parameter space, partly to evaluate the relative merits of different capabilities toward the scientific cases advanced for the E-ELT, while also exploring the instrument requirements and technology readiness of likely components. Three of the studies were of MOS instruments: EAGLE (Cuby et al. 2010), OPTIMOS-EVE (hereafter referred to as EVE, Navarro et al. 2010), and OPTIMOS-DIORAMAS (Le Fèvre et al. 2010), which explored different parts of parameter space in terms of the image quality provided by adaptive optics (AO), number of targets, spectral coverage, spectral resolution, and imaging capability.

The ESO instrument roadmap has identified two first-light instruments for the E-ELT (Ramsay et al. these proceedings): a near-IR imager and a (red-)optical/near-IR integral field unit (IFU) spectrograph, with requirements similar to the Phase A studies for MICADO (R. Davies et al. 2010) and HARMONI (Thatte et al. 2010), respectively. These will exploit the limits of image quality of the E-ELT via high performance AO, but will necessarily be limited in their spatial extent on the sky: i.e. a 1 to 2 arcmin field-of-view imager, and a monolithic (i.e. single-target) IFU.

A HARMONI-like instrument will be well suited to spectroscopy of individual high-$z$ galaxies of interest, and stars in very dense regions (e.g. inner parts of spirals, cluster complexes), but the larger samples needed to explore galaxy evolution, both at high and low redshifts, will require MOS observations. Here we revisit the scientific requirements for a MOS capability on the E-ELT, drawing on elements of the cases advanced for the EAGLE and EVE studies (and excluding cases considered previously, for which spectroscopy with a HARMONI-like instrument would suffice).

The E-ELT has been designed with AO integrated into the telescope, with a large adaptive mirror (M4) and a fast tip-tilt mirror (M5) to correct for the turbulence in the lower layers of the atmosphere via ground-layer adaptive optics (GLAO). Higher-performance AO will be delivered by dedicated modules, or within the instruments themselves. In the context of sources for MOS observations, we divide the cases into two types:

- 'High definition': Observations of tens of channels at fine spatial resolution, with multi-object adaptive optics (MOAO) providing high-performance AO for selected sub-fields in the focal plane (e.g. Rousset et al. 2010).
- 'High multiplex': Integrated-light (or coarsely resolved) observations of >100 objects at the spatial resolution delivered by GLAO.

In what follows, we assume access to a patrol field with an equivalent diameter of 7 arcmin. While the E-ELT has been designed to have excellent image quality across a 10 arcmin (diameter) field-of-view, the infrastructure required for AO guide stars and other sensors limits this to an effective field of ~7 arcmin. Nonetheless, the prospect of a ~40 arcmin$^2$ field on a 40m class telescope is remarkably attractive for spectroscopy. In Sections 2 and 3 we discuss the important MOS cases (and their requirements) for studies of high-redshift galaxies and resolved stellar populations. In Section 4 we combine these into a common set of requirements, and an initial top-level concept is discussed in Section 5.

## 2. THE PHYSICS OF HIGH-REDSHIFT GALAXIES

The cases advanced for studies of high-$z$ galaxies with the E-ELT encompass a broad range of potential targets, from emission-line spectroscopy of the most distant 'first light' galaxies, to spatially-resolved spectroscopy of galaxies at $z$=3-4. Moreover, the sensitivity of the E-ELT will provide our first views of the large populations of high-$z$ dwarf galaxies, investigating their role in galaxy assembly and evolution.

The two most important requirements that determine the scale of a MOS instrument are the field-of-view on the sky from which science targets can be selected (the 'patrol field') and the number of objects/sub-fields that can be observed simultaneously (the 'multiplex'). As already noted, for the former we adopt the maximum field available in the current design of the E-ELT (equivalent to a 7 arcmin diameter), while the source densities of high-$z$ targets depend on the case of interest – we now discuss three of these, and the related case of mapping the intergalactic medium (IGM).

### 2.1. 'First light' galaxies
For a complete picture of the star-formation and quasar activity responsible for reionisation of the early Universe we need an inventory of the first galaxies. In short, analysis of Ly-α emitters and Lyman-break galaxies (LBGs) at different redshifts, via their line emission and luminosity functions. This will constrain when reionisation occurred and identify the galaxies responsible. The first substantial samples of candidate galaxies with $z$≥7 were just appearing at the end of the Phase A instrument studies, with a large number of papers published in this area since then (see review by Dunlop, 2012). Spectroscopic confirmation of photometrically-selected targets at the highest redshifts is rare, and becomes essentially impossible with current facilities. A key issue is the possible trapping of Ly-α by a neutral IGM at $z$>7, potentially making this primary diagnostic feature invisible. Dunlop (2012) flagged the following as some of the key unanswered questions about the highest redshift galaxies:

- Measurement of the faint-end slope of the high-$z$ luminosity function (allowing determination of the origin of the ionising radiation causing the reionisation of the universe).
- Determination of the evolution of the Ly-α luminosity function at $z$>7, giving information about the final stages of reionisation.
- Estimation of the stellar initial mass function, star formation history, metallicity and dust content of galaxies at $z$>7 to allow accurate determination of total stellar masses and specific star-formation rates (i.e. star-formation rates per unit stellar mass already in place)

As noted by Dunlop, one of the key observational goals is therefore to obtain complete spectroscopic follow-up of LBGs, over a wide range of UV luminosity and redshift. The vast collecting area of the E-ELT will enable this huge advance, primarily through spectroscopy of the rest-frame UV features. For the brighter objects, spatially-resolved observations could be used to begin to untangle the physics of star formation at these very early times. For the fainter objects, the flux from a given target may have to be co-added to improve the S/N, but spatially-resolved observations are probably still needed to enable optimal co-addition of the flux along with accurate background subtraction.

### 2.1.1. Requirements

The source densities of galaxies at the largest redshifts and performance estimates for EAGLE spectroscopy were presented by Evans et al. (2010). In brief, the deep imaging from *HST*-WFC3 (e.g. Bouwens et al. 2008, 2010; Oesch et al. 2009) was used to provide an estimate of source densities in different redshift bins. When the success rates of detecting these galaxies (via their rest-frame UV absorption lines) are factored in, we find an expected surface density of 'good' targets of a few tens in the E-ELT field-of-view, i.e. ~0.5-2 arcmin$^{-2}$, depending on the redshift bin (see Evans et al. 2010; Welikala et al. in preparation). Although the high-$z$ galaxies are small, their faintness means that diffraction-limited spectroscopy is not required. The brightest $z$>7 candidate galaxies from the *HST*-WFC3 imaging have half-light radii of ~0.15 arcsec (e.g. Bouwens et al. 2010). Thus, a 'high definition' mode of an ELT MOS will provide multiple spatial elements across the half-light diameters of such systems, enabling us to spatially resolve velocity offsets and stellar populations within these galaxies, providing an important means to constrain their dynamical masses and stellar mass-to-light ratios.

To investigate how well we would recover the sub-components and velocities for different spatial-sampling sizes, we simulated observations of a clumpy galaxy at $z$~7 and $J_{AB}$~27, with a total integration time of 30 hrs; an example of these simulations is shown in Fig. 1. The brightest components are recovered successfully in the case of spatial pixels of 37.5 mas (as considered by EAGLE, and shown in the upper-centre panel in the figure); with spatial pixels of 5 mas, the observation is readout-noise dominated and there is no detection. With MOAO correction and spatial sampling of order 40 mas for the 'high definition' mode of an ELT MOS, it should be possible to resolve individual H$\scriptstyle\rm II$ regions down to scales of ≤200pc in the target galaxy. Since the H$\scriptstyle\rm II$ regions will already have a high contrast due to their emission lines, the gain in sensitivity pushes the telescope's performance into a new regime and will give us unparalleled insight into the dynamical and stellar masses of these early systems.

### 2.2. Spatially-resolved spectroscopy of high-z galaxies

The availability of IFUs on 8-10 m class telescopes has heralded a new era of galaxy studies. We can derive spatially-resolved kinematics and physical properties of distant galaxies at redshifts of up to $z$~3 (e.g., Förster Schreiber et al. 2006 2009; Yang et al. 2008; Law et al. 2009; Lemoine-Busserolle et al. 2010; Contini et al. 2012; Puech et al. 2012). However, above $z$~1 only the most massive, high-surface-brightness systems are within the grasp of current facilities. In addition, most of the samples observed at $z$>1.5-2 have been assembled (necessarily) from a collection of selection criteria. Whether these samples are representative of the early Universe remains uncertain, and our models of galaxy formation and evolution rely on analytic/empirical prescriptions, with inputs such as metallicity, angular momentum, and the spatial distribution of the gas taken from the existing observations.

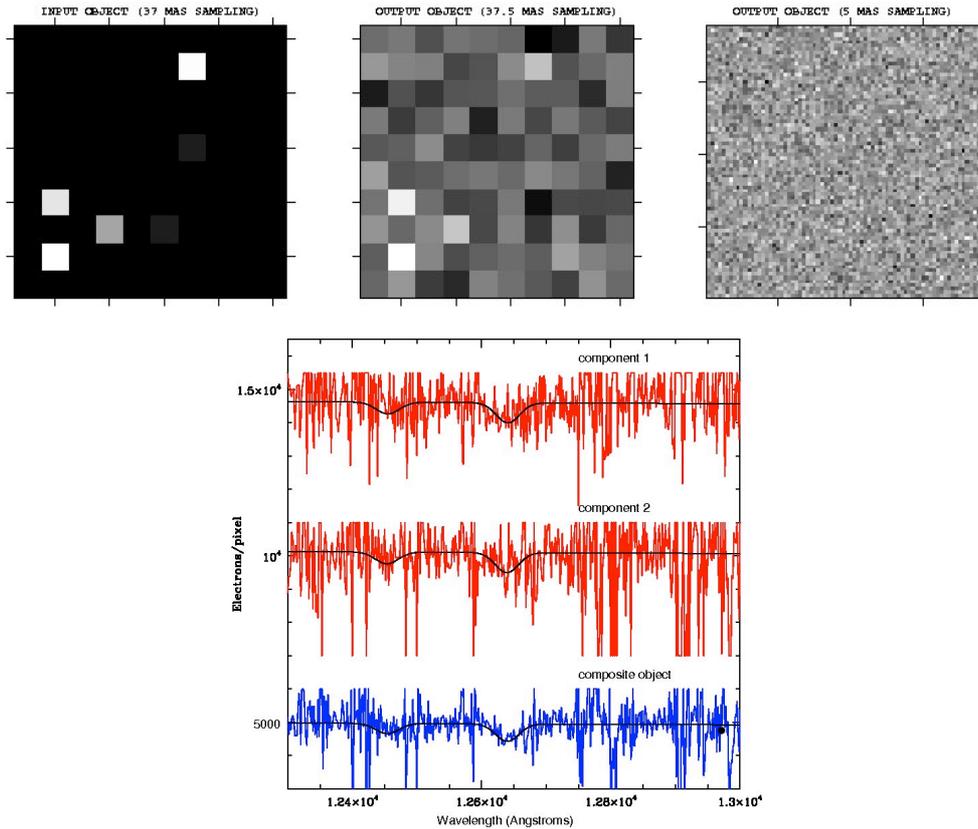

**Fig. 1:** Simulations of a clumpy high-$z$ galaxy at $z\sim7$ (Weliakala et al. in preparation). *Upper panel (from left-to-right):* simulation input with six clumps at $z\sim7$; simulated $H$-band IFU cube for one spectral element at $R = 4000$ (between the OH lines) at a spatial pixel scale of 37.5 mas; the result of the same simulation with spatial sampling matched to the E-ELT diffraction limit, which over-resolves the clumps. *Lower panel:* $J$-band spectra (37.5 mas sampling) of two brightest components, and composite of the four brightest.

The processes by which disk galaxies assembled their mass remains very much debated. For example, the importance of minor vs. major mergers, as well as the accretion of hot vs. cold gas. To make significant progress in our understanding of galaxy formation and evolution requires observations of substantial samples of galaxies, over a large enough volume to rule out field-to-field biases ('cosmic variance') – only then will we be able to test the models rigorously with representative samples. The capabilities of the E-ELT will provide, for the first time, the potential for spatially-resolved observations of an unbiased and unprecedented sample of high-$z$ galaxies.

### 2.2.1. Requirements

The requirements for this case have been discussed by Puech et al. (2008, 2010a). In brief, the goal is to target the rest-frame optical emission lines redshifted into the near-IR. An effective redshift limit to such studies is given by the [OII] emission line leaving the $K$-band (at $z\sim5.6$). However, $K$-band observations will require significantly larger exposure times than at shorter wavelengths because of the increased thermal background (Puech et al. 2010a). If the requirement to observe in the $K$-band is dropped, observations would be limited to $z\sim4$ ($H$-band). A minimum spectral resolving power of $R=4000-5000$ is needed to resolve the brightest OH sky lines and identify emission lines between them; this should also allow resolution of the [OII] doublet in the most distant galaxies.

Preliminary estimates of the number of targets of interest per patrol field (38.5 arcmin²) were estimated to be tens to around a hundred galaxies, taking into account the spectroscopic success rate of measuring emission lines (Evans et al. 2010). Using results of the Design Reference Mission of the E-ELT, a survey of ~160 galaxies, spanning $z$=2-5.6 in three redshift and mass bins, would require ~90 nights of observations (including overheads) with a multiplex of 20 IFUs (see Puech et al. 2010a). A similar survey limited to $z\leq4$ (i.e. omitting the $K$-band) would provide observations of ~240 galaxies in only ~12.5 nights. In contrast, such a sample with a single IFU instrument would require ~250 nights, thus needing 4-5 years of observations (assuming standard operations of sharing between instruments/programmes). Assembling a spectroscopic survey of hundreds of spatially-resolved high-$z$ galaxies remains a strong scientific motivation for the E-ELT, and can only be obtained efficiently with a multi-IFU instrument.

The spatial resolution and corresponding image quality (ensquared/encircled energy) required for the galaxy evolution case was discussed by Puech et al. (2008, 2010a, 2010b). Observations and theory suggest important sub-structures ('clumps') of ~1 kpc in scale in many galaxies (e.g. Elmegreen & Elmegreen, 2005; Bournaud et al. 2007), arguing for sampling on the sky of at least ~40 mas/pixel. A HARMONI-like IFU will almost certainly be used to study individual galaxies (or small samples) on this scale, but the desire to obtain the large-scale surveys discussed above still argues for fine spatial sampling in the case of an ELT MOS. The next relevant spatial scale (in the sense of coarser sampling) for studying distant galaxy structures is the diameter of the target galaxies; this is the scale of interest when studying large-scale motions in which the dynamical nature of galaxies is imprinted (e.g. Puech et al. 2008), arguing for spatial sampling of 50-75 mas/pixel. The size of each IFU is defined by the size of the galaxies on the sky and the need for good sky subtraction (e.g., classical A-B-B-A dithers within the IFU). An IFU size of order 2×2 arcsec$^2$ is sufficient such that only the closest and most massive/luminous galaxies would require specific offset sky measurements.

### 2.3. Sub-L* & dwarf galaxies in the distant Universe

The deepest images from the *HST* have revealed a plethora of intrinsically faint galaxies at $z$=1-3 (Ryan et al. 2007), providing a fairly steep low-mass slope of the UV luminosity function (Reddy & Steidel, 2009); this implies a considerable increase of the dwarf number fraction with redshift. While sub-L* galaxies may contribute to a significant fraction of the evolution in star-formation density, we know little about the populations of dwarf galaxies at high-$z$. Their faint apparent magnitudes ($m_{AB}$~24-27) prevent spectroscopy with current facilities and their morphological appearance is severely affected by cosmological dimming. Beyond understanding the role of sub-L* galaxies in the bigger picture of the evolution of mass and star-formation densities, there are other important questions that can be addressed with a large multiplex MOS (potentially with coverage extending into the near-IR), as illustrated by the following three examples.

### 2.3.1. The impact of low-surface-brightness galaxies (LSBGs) at high-$z$

LSBGs (with $\mu_0(B)$>22 mag arcsec$^{-2}$ or $m_0(R)$>20.7 mag arcsec$^{-2}$; Zhong et al. 2008) represent 9% of the baryonic mass and a third of the H I mass density in the local Universe (Minchin et al. 2004), but their effect on the distant Universe is a complete mystery. LSBGs are gas rich and have average stellar masses comparable to dwarfs, such that they may contribute significantly to the increase of the number density of dwarfs at high-$z$. The gas fraction appears to increase rapidly with $z$ (at a rate of 4% per Gyr from $z$=0 to 2, Rodrigues et al. 2012) and LSBGs might include a significant (dominant?) fraction of the Universal baryonic content at high-$z$. With high-multiplex observations we could investigate their ISM properties (dust, metal abundances, star-formation rates), while complementary IFU observations may provide their ionised gas fractions using inversion of the Kennicutt-Schmidt law.

### 2.3.2. H II galaxies to probe the curvature of the Universe (Λ)

With careful target selection ($W_0(H\beta)$≥50Å, Gaussian profile, continuum dominated by the nebular component), H II galaxies show a remarkably tight correlation (over four dex in luminosity) between the luminosity of recombination lines and the velocity dispersion of the ionised gas, e.g., $L(H\beta)$ vs. $\sigma(H\beta)$. This arises when a starburst fully dominates its host galaxy by the correlation between ionising photons and the turbulence of the ionised gas, with Melnick et al. (2000) showing its validity up to $z$~3. Employing this correlation, Chavez et al. (2012) have used H II galaxies to provide an independent estimate of $H_0$ in the local Universe. Indeed, due to their brightness, H II galaxies at large redshifts (e.g., $z$~1.5-2.4) could provide larger observational samples than those of distant type Ia supernovae, at epochs for which the curvature effect is most pronounced. This could provide a complementary method to evaluate the dark energy equation-of-state at larger redshifts than currently probed by type Ia supernovae (Plionis et al. 2011). Deep exposures will be required to reach S/N~5 in the continuum of H II galaxies with $J_{AB}$~27 (requiring total exposure times of up to ~40 hrs), with a large (>100) multiplex. To reach the continuum level (to avoid contamination by stellar light) and to separate Gaussian Hβ profiles from those that are asymmetric/multiple/rotationally-contaminated argues for $R$≥5000 (with $R$~10000 desirable).

### 2.3.3. The origin of dwarf galaxies

This is a two-fold programme to improve on our knowledge of distant dwarf galaxies. First, the signature of primordial dwarfs can be characterised from the peculiar properties of IZw18-type galaxies, whose spectra show prominent [OIII] λ5007 and Hα equivalent widths (>>1000Å) that severely contaminate the continuum, impacting on the shape of their spectral energy distributions. Deep near-IR imaging may provide targets down to $J_{AB}\geq 29$ and detection of their prominent emission lines will require MOS observations with the E-ELT, enabling determination of the evolution of the number density of primordial galaxies from $z$=0-1.5 (or up to $z$=2.4). Second, it has been claimed that most present-day dwarfs could have a tidal origin (Okazaki & Taniguchi, 2000). This hypothesis can be probed directly at $z$=0.8-2.4, the era at which most of the mergers would have occurred (e.g. Puech et al. 2012). Such follow-up requires accurate redshifts with a large multiplex to probe the 3D locations of dwarfs relative to massive galaxies observed in a major-merger process, with their locations expected to follow the tidal tails that can be modelled at the in-situ merger (e.g. Hammer et al. 2009). Another motivation for such a study would be to assess the number density of satellite galaxies and compare it to the expected number of haloes from the ΛCDM theory, from $z$=2.4 to the present day.

With target magnitudes of $J_{AB}$~26 (27), an exposure time of 17 hrs (in the 'high multiplex', GLAO mode) would provide S/N~8 (3) per continuum pixel. Spectral coverage into the near-IR (up to 1.7μm) is desirable to enable observations of [OII] λ3727 through to Hα out to $z$~1.5 and [OII] λ3727 through to [OIII] λ5007 out to $z$~2.4. Analysis of these lines would investigate the properties of the interstellar medium in these galaxies (dust content, metallicity, abundances, star-formation rates etc). The source densities of high-$z$ dwarfs with $M_{stellar}\leq 3\times 10^9$ $M_\odot$ (i.e, masses comparable to the LMC or lower) with $J_{AB}\leq 26$ (≤27) within the E-ELT patrol field is ~800 (1750), arguing for a large multiplex capability (see Fig. 2).

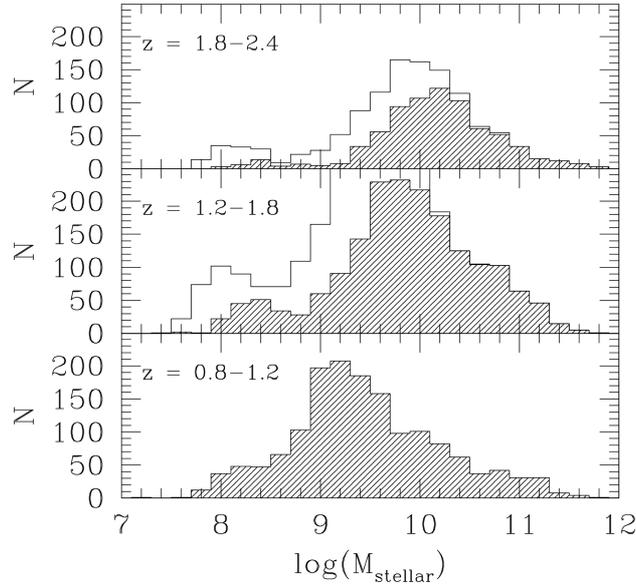

**Fig. 2:** Distribution of stellar masses for galaxies detected by *HST*-WFC3 in the GOODS-S field within an area of 6.8×10 arcmin. Photometric redshifts are derived from Dahlen et al. (2010 and private communication), absolute magnitudes are determined from an interpolation method (Hammer et al. 2001), and stellar masses are determined from the results of Bell et al. (2003). Shaded (and non-shaded in the upper panels) histograms represent galaxies with $J_{AB}\leq 26$ (≤27) with 1460 (3100) dwarfs, with $M_{stellar}\leq 3\times 10^9 M_\odot$.

### 2.4. Tomography of the IGM

The gas in the IGM is revealed by the numerous absorption lines that are seen in the spectra of quasars bluewards of the Ly-α emission line from the quasar. It has been shown that the high-$z$ IGM contains most of the baryons in the Universe and is therefore the baryonic reservoir for galaxy formation. In turn, galaxies emit ionising photons and expel metals and energy through powerful winds which determine the physical state of the gas in the IGM. This interplay of galaxies and gas is central to the field of galaxy formation and happens on scales of the order of 1 Mpc or less (~2 arcmin on the sky at $z$~2.5 using standard cosmological parameters). At larger scales, the gas is in the linear regime and probes large-scale structures. The main goal of this programme is to reconstruct the 3D density field of the IGM at $z$~2.5 to study its topology and to correlate the position of the galaxies with the density peaks.

The Ly-α forest seen in quasar spectra arises from moderate density fluctuations in a warm photo-ionised IGM. The spatial distribution of the IGM is related to the distribution of dark matter and the full density field can be reconstructed using a grid of sight-lines (Pichon et al. 2001; Caucci et al. 2008). A Bayesian inversion method interpolates the structures revealed by the absorption features in the spectra; Fig. 3 shows the level to which the matter distribution could be recovered. One hundred parallel sight-lines were drawn through a 50×50×50 Mpc *N*-body simulation box and synthetic spectra were generated. From analysis of the simulated data it was possible to recover structures over scales of the order of the mean separation of the sight-lines*.*

About 900 randomly distributed targets per square degree would be required to recover the matter distribution, with a spatial sampling of 0.5–2 arcmin at $z$~2. Once the density field is recovered, topological tests can be applied to recover the true characteristics of the density field. To achieve these goals, in addition to high-*z* quasars we also need to target LBGs, thus requiring observations around $\lambda_{obs}$~420 nm (for $z \geq 2.5$). We expect observations of ~20 sources in a 7 arcmin field. A minimum spectral resolving power of $R$=5000 is required (although $R$=10000 would be helpful to avoid metal lines). To reach S/N≥8 at $r$=24.8 (considering the background sources as point-sources), exposures of 8-10 hrs are required per field, thus ~750 hrs to cover one square degree. The wavelength range should cover enough of the Ly-α forest and, in principle, 20 nm is enough. However, one of the main advantages of this method is to reconstruct the IGM at different redshifts, thus arguing for greater wavelength coverage. In addition, observing the objects redwards of the Ly-α emission would allow us to study the LBGs themselves, the metals in the IGM (in the case of quasars), and the quasars themselves if the IR is accessible.

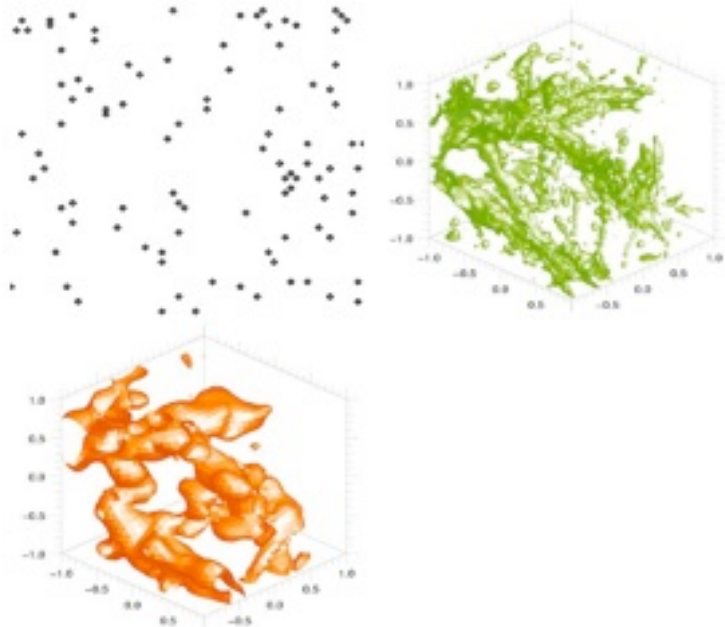

**Fig. 3:** The input simulated density field is shown in the right-hand box (in green). One hundred (randomly-spaced) sight-lines to background sources are drawn through this box (upper-left). The corresponding spectra are used as the input data for the reconstruction, with the reconstructed density field shown in the lower box (in orange).

# 3. SPECTROSCOPY OF RESOLVED STELLAR POPULATIONS

The primary stellar targets for E-ELT spectroscopy are in galaxies beyond the Local Group, such as those in the Sculptor Group (Fig. 4). These are large extended galaxies in which there is a wide range of stellar densities/crowding. A key point to note is that there is already substantial deep imaging available of such galaxies from, e.g., the *HST*; i.e. we already have catalogues of potential targets, but lack the sensitivity with 8-10m telescopes to obtain spectra with adequate S/N. An instrument like HARMONI will be well suited to spectroscopy of stars in individual dense regions in external galaxies (and the Milky Way), but the larger samples needed to explore entire galaxy populations will require MOS observations. The primary objective is to recover estimates of stellar metallicities and radial velocities, which will require both 'high definition' and 'high multiplex' observations (depending on the stellar densities in different regions of the target galaxies), as discussed below.

## 3.1. High multiplex (GLAO) observations

The outer regions of external galaxies can provide some of the most dramatic evidence about their past star-formation histories, via the presence of extended structures, stellar streams etc. The principal diagnostic for estimates of stellar metallicities and radial velocities in this context is the calcium triplet (CaT) in the range of 0.85-0.87μm (e.g. Tolstoy et al. 2001; Koch et al. 2007). At large galactocentric distances we would expect relatively metal-poor populations, but one of the major assets of the CaT is that its three absorption lines are relatively strong, meaning that its relationship to metallicity, [Fe/H], is robust over a large range (e.g. Cole et al. 2004; Carrera et al. 2007), with Starkenburg et al. (2010) providing a new calibration valid to metallicities as low as [Fe/H]≥ −4. Battaglia et al. (2008) demonstrated that, with careful calibration and S/N≥20 Å$^{-1}$, metallicities obtained from the low-resolution mode of FLAMES-Giraffe (*R*~6500) are in agreement with direct measurements from the high-resolution mode (*R*~20000). In practise, *R*≥5000 is sufficient, given adequate S/N (i.e. ≥20). The scientific objective here is observations of hundreds of stars per galaxy but, given the low source densities at large galactocentric distances, high-performance AO is not required and the GLAO correction enabled by M4 and M5 provides acceptable image quality.

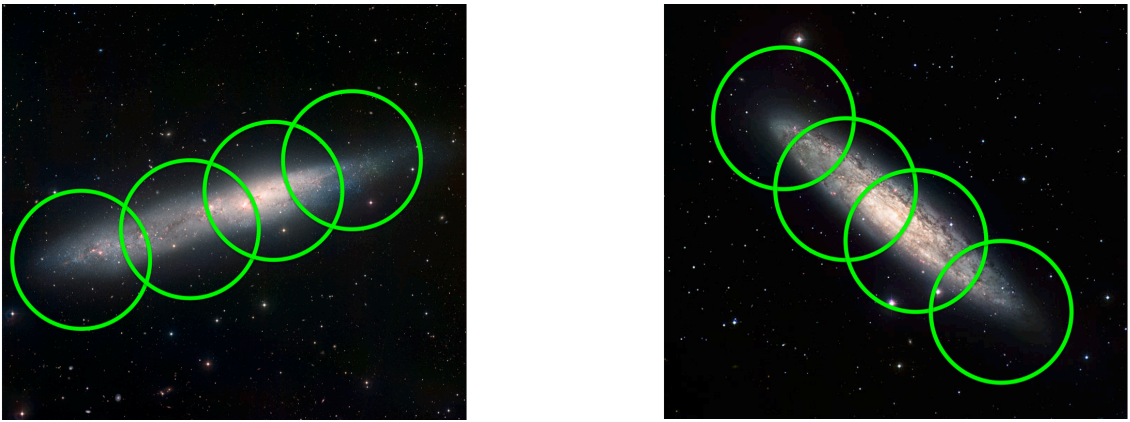

**Fig. 4:** Illustrative 7' diameter pointings in NGC 55 at 1.9 Mpc (*left-hand panel*) and NGC 253 at 3.6 Mpc (*right-hand panel*).

## 3.2. High definition (MOAO) observations

Spatial resolution becomes more of a factor in the denser regions of external galaxies, and therefore makes stronger demands on the AO performance. For instance, the spatial resolution delivered by the *HST* observations with instruments like ACS and WFC3 is more than a factor of three finer than the sampling provided by the GLAO correction – to successfully follow-up such *HST* observations (for example), will require additional correction. While, the CaT is a widely-used spectral feature for studies of stellar populations, the AO correction will be more effective at longer wavelengths, and it is worth considering other diagnostics. Observations in the *J*-band (at *R* of a few thousand) were demonstrated by B. Davies et al. (2010) to provide robust estimates of metallicities of red supergiants (RSGs). The 1.15-1.22μm region includes absorption lines from Mg, Si, Ti, and Fe so, while the lines are not as strong as those in the CaT, they provide direct estimates of metallicity. The potential of this region for extra-galactic observations of both red giant branch (RGB) stars and RSGs with the E-ELT was further investigated by Evans et al. (2011). They concluded that a continuum S/N>50 (per two-pixel resolution element) was required to recover simulated input metallicities to within 0.1 dex, sufficient for many extra-galactic applications.

While the objective is spectroscopy of individual stars (i.e, point-sources), the extended spatial coverage provided by IFUs is an attractive solution – IFUs with fields on the sky of ≥1.0x1.0 arcsec would provide adequate spatial pixels for good background subtraction combined with multiple stars per IFU. Indeed, with potentially tens of stars per IFU, the effective multiplex can be large. We require samples of >>100 stars per galaxy, to adequately sample their different populations and sub-structures (thus requiring multiple E-ELT pointings across each galaxy, e.g. Fig. 4). To assemble ~1000 stars per target galaxy (within a few pointings) suggests a multiplex in the range of 10 to 20 IFUs. The driving performance requirement on the AO system for the 'high definition' cases is the encircled energy at longer wavelengths (in the $H$-band) for the high-$z$ targets. This is already demanding in the context of a wide-field MOAO system, so the (necessarily) pragmatic approach for the stellar case is to scale these performances to shorter wavelengths (e.g. the performances for different asterisms in Table 3 of Evans et al. 2011).

### 3.3. Performance estimates

The simulations presented by Evans et al. (2011) were of MOAO observations of both the CaT and in the $J$-band, employing the web-based simulation tool developed by Puech et al. (2008). Given the diminished AO correction at shorter wavelengths, we have simulated new CaT observations with coarser spatial sampling from GLAO (Section 3.1) using a version of the Puech et al. tool for simulations of EVE performances. To enable a direct comparison with the previous simulations, we adopted similar parameters as Evans et al. (2011): a total throughput of the telescope of 80%, an instrumental throughput (including the detectors) of 35%, a low read-out noise of 2e$^-$/pixel, and an exposure time of 20×1800s. Other inputs to the calculations were a spatial sampling of 0.3 arcsec (of a 0.9 arcsec aperture), $R$=5000, and one of the $I$-band GLAO point-spread functions (PSFs) from Neichel et al. (2008). The PSF included a ring of nine natural guide stars at a diameter of 7 arcmin – the configuration of guide stars (both natural and laser) will be somewhat different to this, but this PSF serves as a first-order test of the likely performances in the GLAO case (excluding the known limitations in the simulations).

In parallel, on-sky tests of the $J$-band methods are underway via X-Shooter observations of RSGs in the Magellanic Clouds. Once the additional problems of real observations are taken into account (e.g. telluric subtraction), analysis of these data suggests that the required S/N for good metallicity estimates is more towards 100 (Davies et al. in preparation). Taking the results from Table 4 of Evans et al. (2011) a S/N~100 is achieved for $J$~22.5 depending on the AO asterism and observing conditions/zenith distance; this sensitivity estimate also takes into account the known limitations in the simulated MOAO performances.

From ten simulation runs to calculate GLAO spectra of the CaT, we find a continuum S/N=21±3 (per pixel) for $I$=23.5, sufficient to recover the metallicity (cf. Battaglia et al. 2008). As with the MOAO PSFs, the simulated GLAO PSFs are also likely somewhat 'optimistic' in terms of the delivered performance. The intrinsic $I-J$ colours for RGB stars are ~0.75 mag so, to first order, the GLAO observations of the CaT and the MOAO observations in the $J$-band provide sufficient S/N to recover the metallicity of a given RGB star – i.e. they are roughly competitive for a given target (ignoring the effects of extinction and crowding for now).

MOS spectroscopy on the E-ELT will provide the capability to determine stellar metallicities out to Mpc distances (for stars near the tip of the RGB) and, in the case of the AO-corrected $J$-band observations, out to tens of Mpc for RSGs. We propose that GLAO observations of the CaT will be sufficient for observations in the sparse regions of external galaxies, and that MOAO observations in the $J$-band are well suited for investigation of the main bodies of the target galaxies. Provision of both of these modes is highly complementary, probing different regions and populations (thus providing good sampling of each spatial and kinematic feature). This combination of modes will give a complete view of each galaxy which is required to confront models of star-formation histories and past interactions. This provides a relatively 'clean' split in the required wavelength coverage for the two modes at ~1μm although, pending more detailed results of the $J$-band methods, extension of the wavelength coverage of the IFUs to 0.8μm is retained as a goal.

### 3.4. High-resolution stellar spectroscopy

There are several science cases which require a greater spectral resolving power ($R$~20000) for more detailed studies of stellar abundances; these also require a large multiplex and do not make strong demands on the angular resolution, thus can be achieved with GLAO.

### 3.4.1. Lithium in Local Group galaxies

We are now discovering the existence of metal-poor populations in nearby galaxies. Analogous to the metal-poor populations found in the Milky Way, these stars are the fossil records of the early stages of the formation of their host systems. Spite & Spite (1982) made the ground-breaking discovery that metal-poor stars at the main-sequence turn-off (MSTO) in the Milky Way display a constant Li abundance, whatever their metallicity or effective temperature. This constant Li abundance is usually called the 'Spite Plateau'. The most straightforward interpretation of this result was that the observed Li was primordial, i.e. produced during the first three minutes of the Universe. In those earliest moments only nuclei of deuterium, two stable He isotopes ($^3$He and $^4$He), and $^7$Li were synthesised. Their abundances depend on the baryon/photon ratio, thus on the baryonic density of the Universe. In principle, the Spite Plateau allows us to determine the baryon/photon ratio, which can not be deduced from first principles.

This interpretation of the Spite Plateau is seriously challenged by the measurement of the baryonic density, with unprecedented precision, from the fluctuations of the cosmic-microwave background by the *Wilkinson Microwave Anisotropy Probe (WMAP)* satellite (Spergel et al. 2007). However, we have indications that the plateau is Universal. For instance, in ω Cen, generally considered to be the nucleus of a disrupted satellite galaxy, Monaco et al. (2010) found the Spite Plateau. It is of paramount importance to verify if this is the case in other Local Group galaxies which possess metal-poor populations, but this requires spectroscopy (at $R \geq 20000$) of stars at the MSTO which, in the Sagittarius Dwarf is at $I \sim 21.5$. These observations are beyond the grasp of the 8-10m class telescopes, but will be accessible with the E-ELT.

### 3.4.2. Exoplanets in Local Group galaxies

Although more than 600 exoplanets are known to exist, they are all confined to a few parsec from the Sun. We have no idea of the dependence of the environment on planet formation and evolution. Fundamentally, the question is whether exoplanets exist also in other galaxies, and whether they are similar to those found in the Milky Way? These questions can be addressed if we can measure precise radial velocities ($\Delta v \leq 10 ms^{-1}$) for giant stars in Local Group galaxies; sufficient to detect 'hot Jupiters'. The Sagittarius Dwarf and the Magellanic Clouds have suitable targets with magnitudes in the range of 18-20 which will require the sensitivity of the E-ELT for such observations. It will be necessary to monitor the radial velocities of many stars in such a search, hence the importance of MOS observations.

## 4. ELT-MOS: COMBINED REQUIREMENTS

Satisfying the two top-level types of observations for a MOS on the E-ELT ('high definition' and 'high multiplex') requires two set of requirements, as summarised in Tables 1 and 2. The requirements for the IGM case (Section 2.4) are summarised in Table 3; these are slightly different as they require optical IFU observations, but without strong requirements on the spatial resolution (i.e. GLAO is sufficient).

Table 1: 'High Definition' Requirements (MOAO)

| Parameter | Essential | Goal |
|---|---|---|
| IFU field-of-view | ≥ 1.5" x 1.5" | ≥ 2.0" x 2.0" |
| Multiplex | ≥ 10 IFUs | 30 IFUs |
| Spatial pixel scale | 50-60 mas | 40 mas |
| Encircled energy (*H*-band) | ≥ 30% in 2x2 spatial pixels | ≥ 40% in 2x2 spatial pixels |
| Spectral resolving power (*R*) | 8000 | … |
| λ-coverage (not simultaneous) | 1.0 – 1.8 μm | 0.8 – 2.45 μm |

**Table 2: 'High Muitiplex' Requirements (GLAO)**

| Parameter | Essential | Goal |
|---|---|---|
| On-sky aperture | 0.9" | … |
| Multiplex | ≥ 200 | ≥ 400 |
| Spatial pixel scale | ≤ 0.9" | … |
| Spectral resolving power ($R$) | ≥ 5000 & 20000 | … |
| Simultaneous λ-coverage | 0.40 – 1.0 μm | 0.37 – 1.35 μm<br>Possible ext. to 1.70μm (TBC) |

**Table 3: Requirements for IGM Case**

| Parameter | Essential | Goal |
|---|---|---|
| IFU field-of-view | ≥ 2.0" x 2.0" | … |
| Multiplex | ≥ 10 IFUs | ≥ 30 IFUs |
| Spatial pixel scale | 0.3" | … |
| Spatial resolution | GLAO | … |
| Spectral resolving power ($R$) | 5000 | 10000 |
| Simultaneous λ-coverage | 0.40 – 1.0 μm | 0.37 – 1.0 μm |

## 5. CONCEPTUAL DESIGNS: OPTIONS

The space available at the Nasmyth platform will depend on the final instrument interface (and the interface for pick-off of the laser guide stars). However, an important asset of the current f-ratio of the telescope is that it delivers a relatively large focal plane – the plate scale leads to a focal plane with a diameter well in excess of 1m. Thus, even with ~200 targets it will not be prohibitively crowded.

The 'high definition' (MOAO-assisted) observations are satisfied by IFUs in the near-IR, while the 'high multiplex' (GLAO) observations are primarily at visible wavelengths. This suggests independent modular spectrographs (or cameras) for the near-IR and visible wavelengths. As noted though, there are also likely to be cases seeking a high multiplex (>>20 objects/field) for near-IR, integrated-light (i.e. GLAO) spectroscopy. Taking advantage of the previous functional analyses done for EAGLE and EVE it should be possible to converge to a compact and modular instrument design that will meet the science requirements defined in Section 4. The overall instrument architecture for an ELT MOS is depicted in Fig 5, which comprises the following sub-systems:

- *Science Object Pick-Off Sub-System (SOPOS):* Refers to the front-end of the instrument and is responsible for selecting science sub-fields/targets and distributing these to the correct spectrographs. This sub-system could be based on a pick-and-place positioner to configure pick-off mirrors/fibre buttons in the focal plane (Fig. 6).
- *Near-Infrared Integral Field Units (NIFUs):* The incoming light from the science objects will be reflected to the front-end receiving optics of the spectrograph. The front-end module will correct the incoming light to negate the effects of atmospheric turbulence and deliver the light to the IFUs of the spectrograph. The spectra from two IFU channels will then be combined onto a single camera detector. The NIFUs provide spectroscopy of one of the *YJ, H, K* bands, at $R$~8000.
- *Visible Spectrographs:* Fibre-fed spectrographs providing simultaneous coverage from 0.40-1.0 μm at $R$~5000. They will have a secondary mode which provides $R$~20000 spectroscopy of selected regions.
- *Adaptive Optics Sub-System (AOS):* The AOS receives data from the wavefront sensors for the natural and laser guide stars (NGS and LGS, respectively), measures the appropriate parameters, and then corrects the wavefront across each of the science sub-fields selected by the free-standing pick-off mirrors. In addition, the AOS corrects the wavefront across the full Nasmyth focal plane using M4 and M5 of the telescope (in effect, wide-field GLAO correction).

- *Laser Guide Star Pick-Off Sub-System (LGS POS):* Picks-off the light from the LGS and delivers it to the appropriate wavefront sensors for real-time analysis of the atmospheric perturbations.
- *Natural Guide Star Pick-Off Sub-System (NGS POS):* Picks-off the light from NGS within the patrol field and delivers it to the appropriate wavefront sensors for real-time analysis of the atmospheric perturbations. The NGS will also be used to improve telescope guiding via correction factors sent to the telescope control system.
- *Instrument Control Sub-System:* Provides for the end-to-end control of the instrument, from user and engineering support, to execution of observations and control of the instrument functions.
- *Science Data Sub-System:* Provides for the assembly, handling, reduction, and temporary storage of the science data. Also responsible for real-time/off-line reduction and quality control of science and calibration data.
- *Instrument Core Sub-System:* The instrument core includes the static structure, field de-rotator, focal plate, ancillary platform and cable wrap, electronics racks, transport, maintenance tools and facilities.

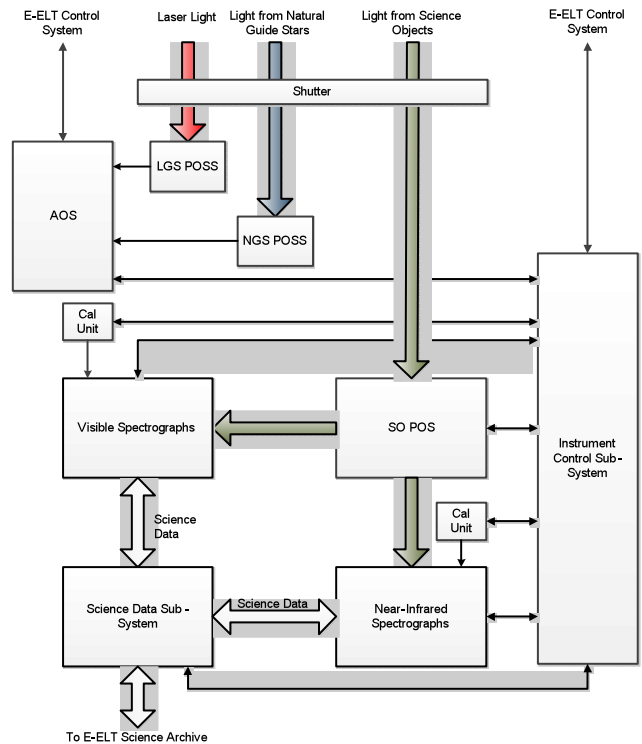

**Fig. 5:** Example instrument architecture for an ELT MOS.

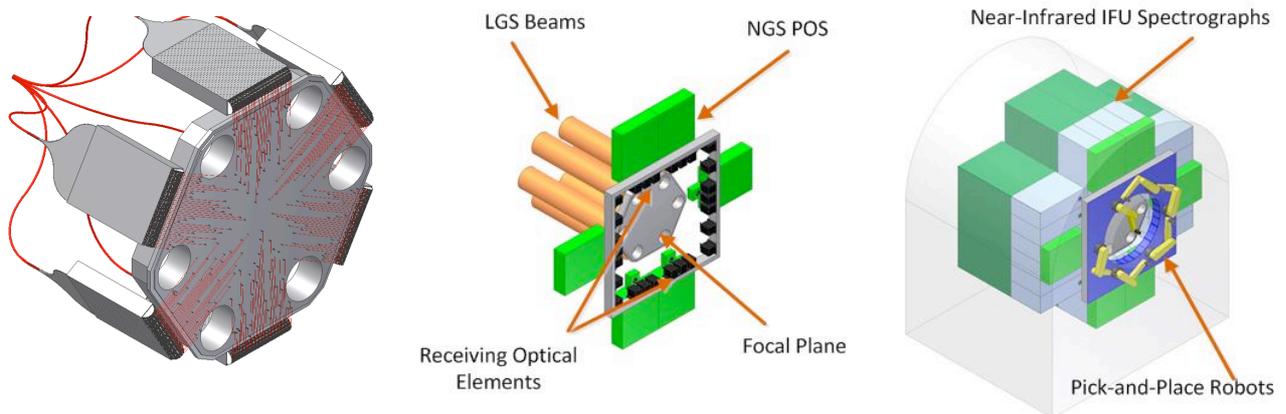

**Fig. 6:** *Left-hand panel:* ELT MOS target selection concept; *Right-hand panels:* ELT MOS sub-system model.

A conceptual model of the ELT-MOS sub-systems is shown in Fig. 6. The visible spectrographs will be mounted nearby on the Nasmyth platform, fed with the fibres which are routed through the centre of the instrument. It might also be possible to avoid having the NIFUs as part of the rotating part of the instrument, by mounting the receiving optical elements on a static structure and using, for example, small rotating pick-off mirrors.

### 5.1. Combined high-multiplex/high-definition observations

The two AO modes discussed above are complementary in the sense that the regions 'between' the MOAO sub-fields will have enhanced image quality roughly equivalent to that from GLAO[1]. If the spectrographs for both modes are independent and the method of target selection enables parallel observations of different samples of targets, this could be an effective means to boost the operational efficiency of the E-ELT. Multi-wavelength surveys have (necessarily) concentrated on a small number of deep fields for studies of distant galaxies, e.g., the *Hubble* Ultra-Deep Field, the *Chandra* Deep Field South, the COSMOS field, etc. Given the huge investment of resources in these existing fields, the E-ELT will certainly be used to observe high-*z* galaxies located within them. Moreover, if new regions are observed to a comparable depth (and breadth of wavelengths), they will still be relatively limited in number. Thus, many of the high-*z* targets envisaged for the cases mentioned in Section 2 are in the same regions of the sky. The same is also true for the stellar populations cases, where there are common galaxies of interest.

To visualise this situation, a (modified) schematic of MOAO is shown in the left-hand panel of Fig. 7. An array of laser and natural guide stars will be used to map the turbulence along the path of each IFU target (e.g. the galaxies in the red boxes in the figure). Additional stars and/or galaxies between these sources could then be picked-off and relayed to a GLAO spectrograph (e.g. the green stars in the figure). An example case for stellar populations in NGC 55 is shown in the right-hand panel of Fig. 7. By virtue of pushing toward the sensitivity limits of the E-ELT, the cases advanced in Sections 2 and 3 will require long exposure times (e.g. 1800s or more per exposure). Thus, in many of the potential cases of parallel observations, there would likely be no need to move to a different target field more quickly with one set of targets than the other. Moreover, targets could be selected to ensure optimal matching of exposure times, still leading to an overall gain in efficiency for the E-ELT compared to separate programmes with both modes.

One of the obstacles to parallel observations could be different requirements for sky subtraction between optical and near-IR targets. Temporal variations in the sky emission-lines in IFUs can be mitigated using the approach of Davies (2007), and accurate near-IR sky subtraction is a field of active investigation, following-on from EVE (e.g. Rodrigues et al. 2010) and for the VLT-MOONS design study (Rodrigues et al. these proceedings). Moreover, if A-B-B-A offsets were necessary for sky subtraction of some IFU observations, different targets for the GLAO spectrograph could be observed in the two positions (with the targets selected for correspondingly shallower observations).

### 6. SUMMARY

We have revisited some of the highlight scientific cases advanced for the E-ELT and derived a set of requirements for an ELT MOS, which combines MOAO-corrected, near-IR spectroscopy, with GLAO-corrected optical spectroscopy (with the latter perhaps extending slightly into the near-IR). Such an ELT MOS will be versatile in its operations and its applications (with many other cases beyond those mentioned here). The inclusion of capabilities to observe with the GLAO provided by the telescope, or MOAO within the instrument, means that the MOS will be available for use in a wide range of atmospheric conditions at the E-ELT. The concept of parallel observations exploiting both spatial scales for some target fields appears attractive in terms of the overall efficiency of the E-ELT – the operational consequences of this for the design of the instrument and the requirements on sky subtraction are an area of ongoing study.

*Acknowledgements*: We thank Colin Cunningham for his input in the early stages of this work.

---

[1] One of the DM technologies considered by the EAGLE study has sufficient stroke and temporal bandwidth such that M4 and M5 can be used to (effectively) deliver a GLAO correction across the full patrol field.

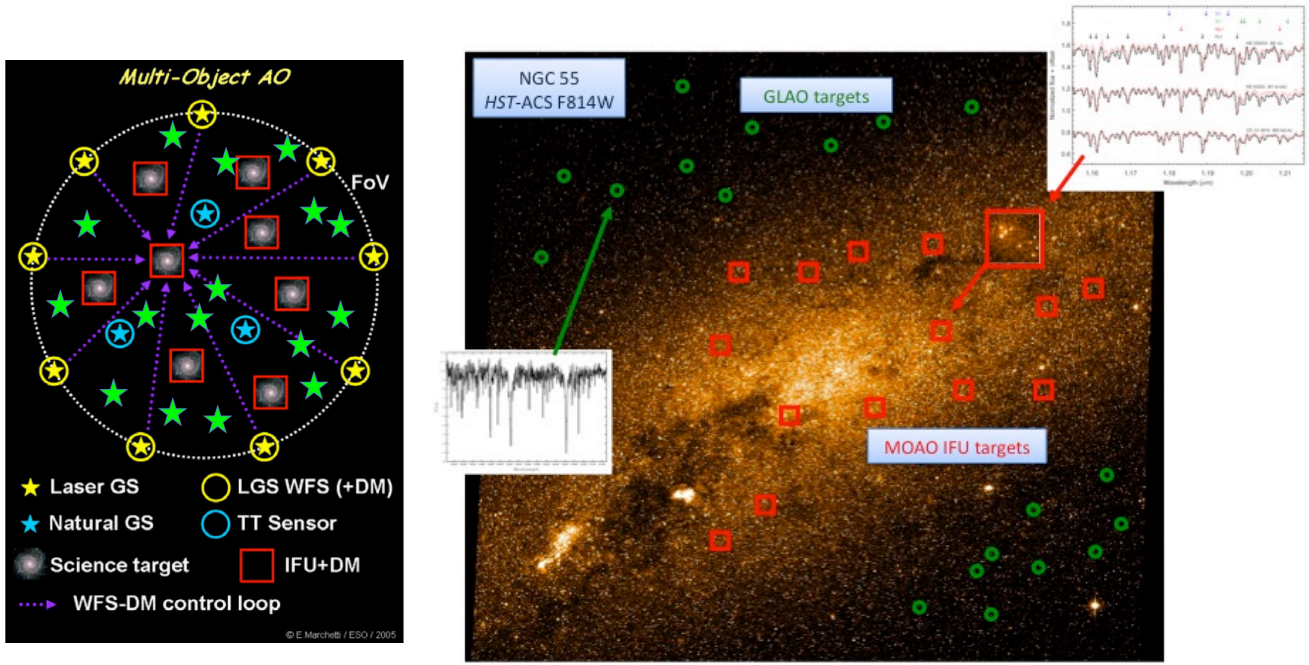

**Fig. 7:** *Left-hand panel:* Schematic of MOAO IFU observations (red squares), with example targets for GLAO spectroscopy elsewhere in the focal plane (green stars). *Right-hand panel:* An example scenario for observations of the stellar populations in NGC 55.